\documentclass[twocolumn,twoside,slac_two]{revtex4}
\pdfoutput=1
\usepackage{graphicx}
\usepackage{bm}
\usepackage{fancyhdr}
\usepackage{natbib}
\setcitestyle{numbers,square}

\begin{document}

\title{Pion Production Cross-section Measurements in~p+C~Collisions at~the~CERN SPS for~Understanding Extensive Air Showers}

%

\author{Marek Szuba for the NA61/SHINE Collaboration}
\affiliation{Karlsruhe Institute of Technology, Germany}

\begin{abstract}
  An important approach to studying high-energy cosmic rays is the investigation
  of the properties of extensive air showers; however, the lateral distribution
  of particles in simulations of such showers strongly depends on the applied
  model of low-energy hadronic interactions. It has been shown that many
  constraints to be applied to these models can be obtained by studying
  identified-particle spectra from accelerator collisions, in the energy range
  of the CERN Super Proton Synchrotron. Here we present measurements of the
  pion production cross-section obtained by the NA61/SHINE experiment at the
  SPS, in proton-carbon collisions at the beam energy of 31 GeV from the year
  2007. Further analyses of identified-particle yields in SHINE, in particular
  with a pion beam, are in preparation.
\end{abstract}

\maketitle

\thispagestyle{fancy}

\section{\label{sec:introduction}Introduction}
  The most common way employed nowadays to study high-energy cosmic rays is to examine
  the properties of extensive air showers (EAS) they induce in the atmosphere, using
  several different observables and detection techniques. The latter include measurements
  of shower size and composition at ground level using surface arrays of scintillation or
  Cherenkov-light detectors, observation of energy losses of shower particles as they
  traverse the atmosphere using fluorescence detectors, detection of radio emission from
  the shower using appropriate antennae, and others. Examples of experiments measuring
  EAS include the surface arrays KASCADE and KASCADE-Grande as well as the hybrid
  surface-and-fluorescence Pierre Auger Observatory~\cite{Antoni:2003gd, Navarra:2004hi,
  Abraham:2004dt}.

  Unfortunately, determination of properties of the initial particle from such
  observables strongly depends on models, especially those of hadronic interactions
  occurring during shower development. Many such models exist yet they frequently fail
  to consistently and accurately reproduce experimental data, for instance
  underestimating the yield of shower muons at ground level or showing non-smooth
  transition from low- to high-energy hadronic interactions~\cite{Bluemer:2009zf,
  Meurer:2005dt, Antoni:2001pw, Abraham:2009ds}.

  In light of the above, additional data is required in order to appropriately tune models
  of hadronic interactions in EAS. Such data can be provided by accelerator experiments
  observing collisions of hadrons such as protons or pions with light ions such as carbon
  nuclei. In particular, it has been demonstrated that the energy range of the Super Proton
  Synchrotron (SPS) at CERN makes it highly suitable for reproducing final hadronic
  interactions\footnote{That is, interactions producing hadrons which do not interact
  further but decay into leptons instead.} in showers of energies studied by KASCADE,
  KASCADE-Grande and the Pierre Auger Observatory~\cite{Meurer:2005dt}.

  This article presents the first results by the NA61/SHINE experiment at the SPS
  obtained for the purpose of measuring the particle yield in the long-baseline
  neutrino experiment T2K and tuning EAS models --- pion spectra from \textit{p+C}
  collisions at~31~GeV.

\section{\label{sec:shine}The NA61/SHINE Experiment}
\begin{figure*}[t]
    \begin{center}
      \includegraphics[width=0.32\textwidth]{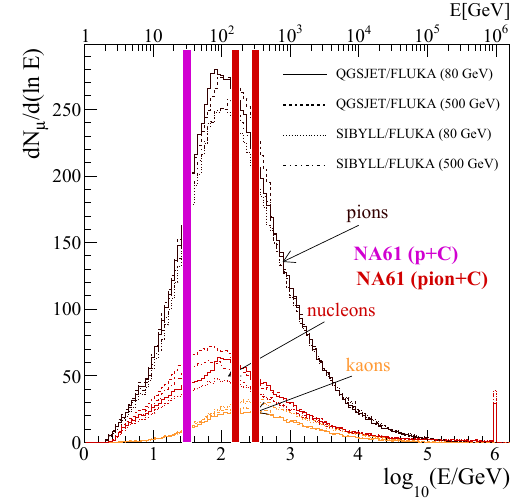}
      \includegraphics[width=0.32\textwidth]{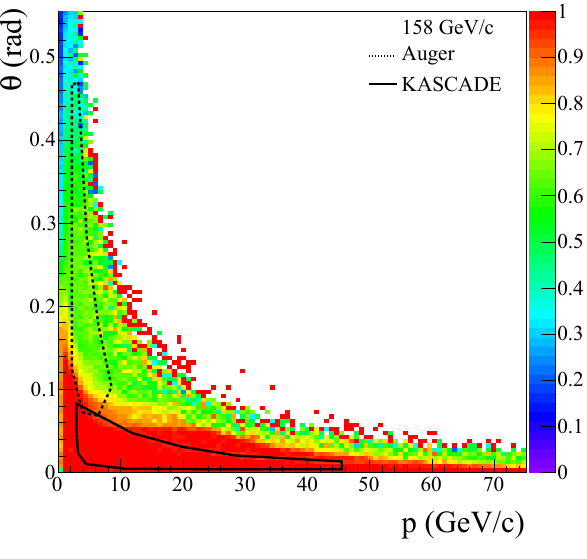}
      \includegraphics[width=0.32\textwidth]{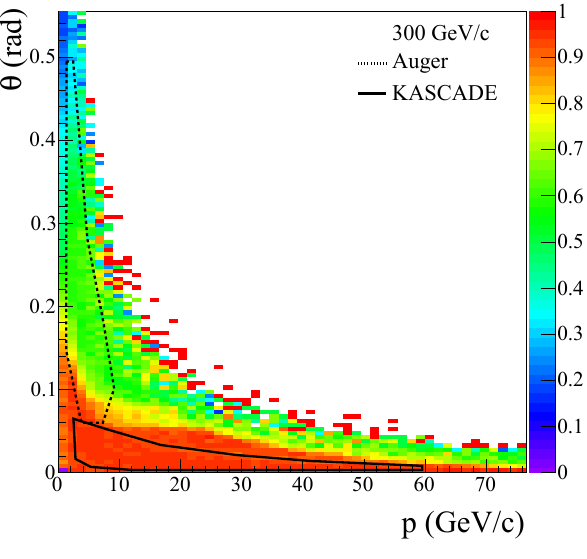}
    \end{center}
    \caption{\textbf{Left}: Energy distributions, simulated using several models,
      of the ``grandfather'' particles in extensive air showers with $E_{0}$ = $10^{15}$~eV,
      with vertical lines indicating the beam energy in relevant NA61/SHINE runs.
      \textbf{Middle and right}: coverage of NA61 in pion--carbon collisions at 158
      (middle) and 300 (right)~GeV \textit{vs} that of KASCADE and Auger, with contours
      indicating the 66-percent level for each~\cite{Maris:2009x1}.}
    \label{fig:na61coverage}
  \end{figure*}

  NA61/SHINE is an experiment at the CERN SPS using the upgraded NA49 hadron
  spectrometer to accomplish a number of physics goals. In addition to providing reference
  data for cosmic-ray experiments it shall also provide model-tuning information for
  the neutrino experiment T2K, produce for the first time in the SPS energy range
  large-statistics proton-proton data sets for high-$p_{T}$ studies and, last but not least,
  perform a comprehensive energy and system-size scan in search for the QCD critical point.
  Its large acceptance (around 50~\% for $p_{T} \le$ 2.5~GeV/c), high momentum resolution
  ($\sigma(p)/p^{2} \approx 10^{-4}~(GeV/c)^{-1}$) and tracking efficiency (over 95~\%),
  and excellent particle-identification capabilities ($\sigma(\frac{\mathrm{d}E}{\mathrm{d}x})
  / \frac{\mathrm{d}E}{\mathrm{d}x} \approx 4~\%, \sigma(t_{ToF}) \approx 100~ps$) make it
  an excellent tool for investigating hadron spectra. Moreover, its kinematic range covers well
  that of KASCADE, KASCADE-Grande and the Pierre Auger Observatory (see Figure~\ref{fig:na61coverage}).

\begin{figure*}[t]
    \begin{center}
      \includegraphics[width=\textwidth]{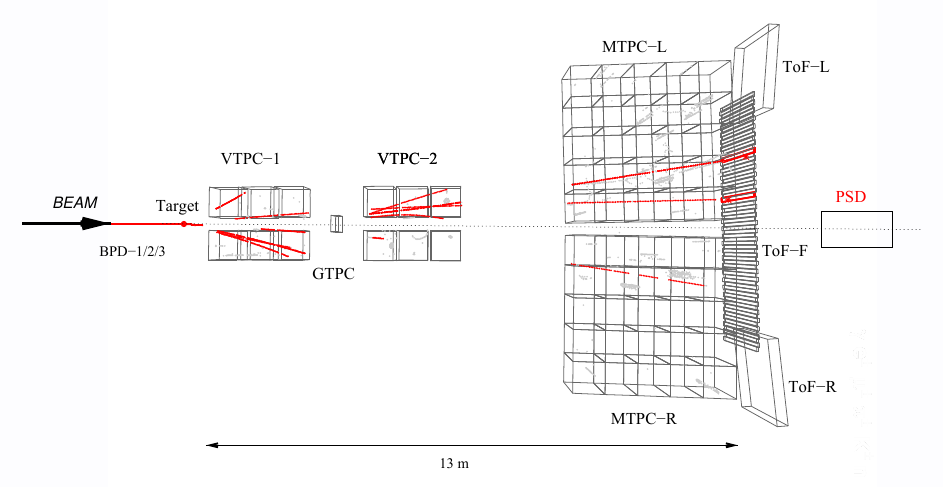}
    \end{center}
    \caption{A view of an NA61/SHINE \textit{p+C} collision, superimposed on the layout
      of the apparatus.}
    \label{fig:na61diagram}
  \end{figure*}

  The following are the main features of the NA61 detector as shown in
  Figure~\ref{fig:na61diagram}~\cite{Afanasev:1999iu, Antoniou:2006mh}:
  \begin{itemize}
    \item tracking plus momentum, charge and $\mathrm{d}E/\mathrm{d}x$ measurement
      with five Time-Projection Chambers;
    \item three Time-of-Flight walls for additional identification information;
    \item high-precision downstream Projectile Spectator Detector;
    \item a number of beam and triggering detectors.
  \end{itemize}

  For the purpose of cosmic-ray studies, SHINE acquired in the years 2007 and 2009 5.4 million
  \textit{p+C} events at~31~GeV, 3.6 million \textit{$\pi^{-}$+C} events at 158~GeV and 4.7 million
  \textit{$\pi^{-}$+C} events at 350~GeV. The analysis of these data sets is currently in progress.
  The results presented here have been produced from the 0.6 million \textit{p+C}-at-31~GeV events
  registered during the 2007 pilot run.

\section{\label{sec:method}The Method}
  The pion spectra presented here have been obtained using three independent
  analysis techniques:
  \begin{itemize}
    \item The $h^{-}$ method, in which all negative hadrons produced in a collision
      are assumed to be pions and the contribution of other species is corrected for using
      simulations. Pros: simple, high statistics. Cons: stronger model dependence, doesn't work
      for positive pions;
    \item $\mathrm{d}E/\mathrm{d}x$ identification of $\pi^{\pm}$. Pros: explicit identification,
      still high statistics thanks to NA61 design. Cons: only works in the momentum regions where
      Bethe-Bloch bands do not overlap;
    \item $\mathrm{d}E/\mathrm{d}x$-plus-ToF identification of $\pi^{\pm}$. Pros: explicit
      identification over a wide momentum range. Cons: limited acceptance.
  \end{itemize}
  For all three approaches, particles passing all the cuts are divided into $(p,~\theta)$
  bins, where $p$ is the total momentum and $\theta$ is the polar angle, to account for changing
  detection and identification properties.

  Last but not least SHINE has estimated the systematic uncertainty of results obtained
  from all three analyses. At present this has been found to be less than on equal to 20~\%;
  we are now working on reducing the systematic uncertainty further.

\section{\label{sec:results}Results}
\begin{figure*}[t]
    \begin{center}
      \includegraphics[width=0.87\textwidth]{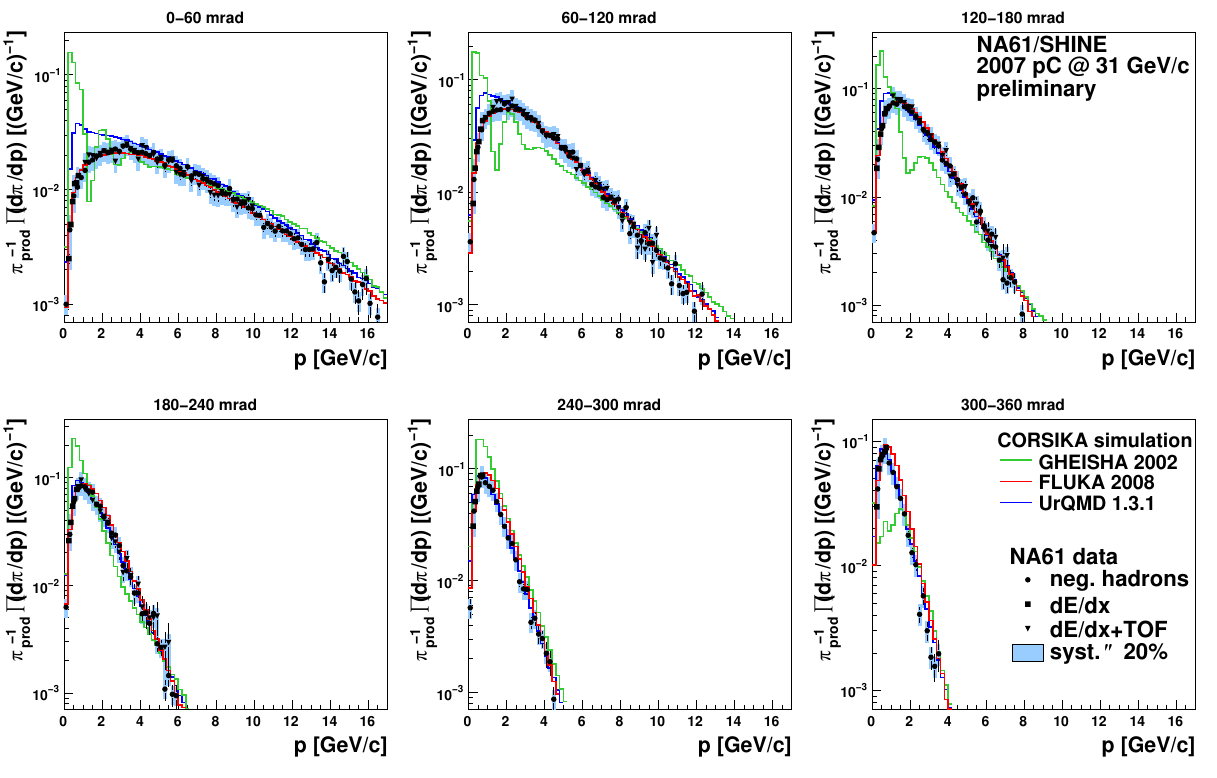}
    \end{center}
    \caption{Momentum dependence of $\pi^{-}$ production cross-section in \textit{p+C}
      collisions at 31~GeV/c. Circles: $h^{-}$ analysis; squares: pions identified
      using $\mathrm{d}E/\mathrm{d}x$; triangles: pions identified using combined
      $\mathrm{d}E/\mathrm{d}x$ and ToF information. Vertical bars and boxes indicate
      statistical and systematic uncertainties, respectively. Lines indicate model
      calculations.}
    \label{fig:xsecMinus}
  \end{figure*}

\begin{figure*}[t]
    \begin{center}
      \includegraphics[width=0.58\textwidth]{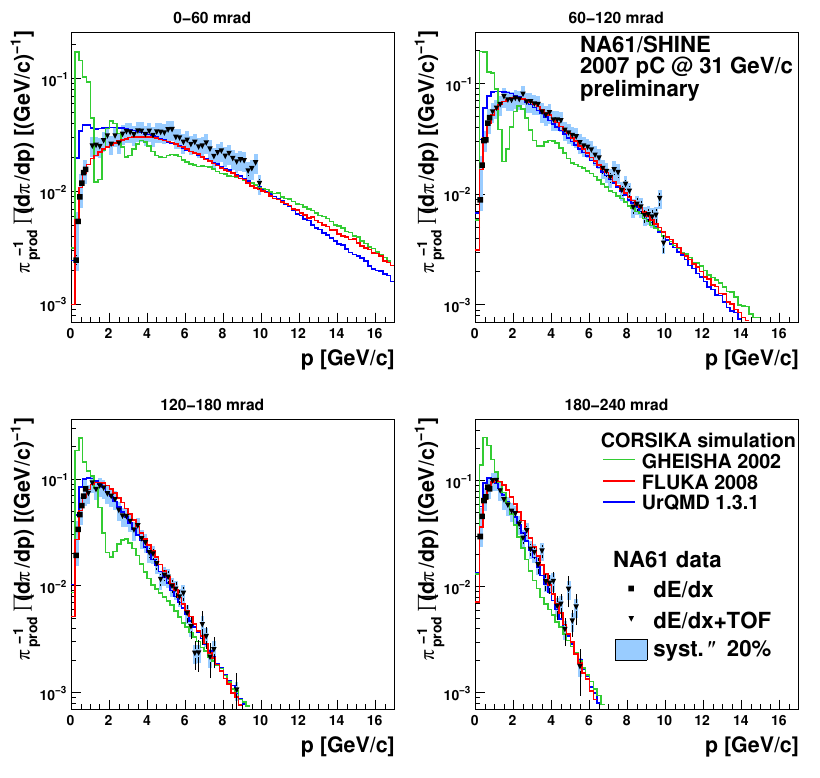}
    \end{center}
    \caption{Momentum dependence of $\pi^{+}$ production cross-section in \textit{p+C}
      collisions at 31~GeV/c. Squares: pions identified using $\mathrm{d}E/\mathrm{d}x$;
      triangles: pions identified using combined $\mathrm{d}E/\mathrm{d}x$ and ToF
      information. Vertical bars and boxes indicate statistical and systematic
      uncertainties, respectively. Lines indicate model calculations.}
    \label{fig:xsecPlus}
  \end{figure*}

  Figures~\ref{fig:xsecMinus} and~\ref{fig:xsecPlus} show the production cross-section
  ($\sigma_{prod} = \sigma_{inel} - \sigma_{qel}$, where $\sigma_{inel}$ and $\sigma_{qel}$
  are inelastic and quasi-elastic cross-section, respectively) of negative and positive
  pions in different $\theta$ bins, compared to air-shower simulations based on the
  CORSIKA package, using three different interaction models: GHEISHA,
  FLUKA and UrQMD~\cite{Heck:1998aa, Fesefeldt:1985yw, Fasso:2000hd, Bass:1998ca}. Please
  see the summary for a discussion of these results.

\section{\label{sec:summary}Summary and Outlook}
  NA61/SHINE has produced its first results relevant to tuning models of hadron
  production in extensive air showers: preliminary $\pi^{\pm}$ spectra from \textit{p+C}
  collisions at 31~GeV. The spectra were obtained using three different methods,
  with very good agreement observed between them. Systematic uncertainties are at the moment
  no greater than 20~\%, with work ongoing to reduce them further. Last but not least,
  preliminary comparisons with simulations show good agreement with FLUKA for polar angles
  below 180~mrad and with UrQMD above that threshold.

  We are now working on finalising and publishing these results, as well getting ready
  to analyse the large-statistics \textit{p+C} and \textit{$\pi$+C} runs from 2009.

\bigskip 
\begin{acknowledgments}
  This work has been supported by  
  the Hungarian Scientific Research Fund (OTKA 68506),
  the Polish Ministry of Science and Higher Education (N N202 3956 33),
  the Federal Agency of Education of the Ministry of Education and Science
  of the Russian Federation (grant RNP 2.2.2.2.1547) and
  the Russian Foundation for Basic Research (grants 08-02-00018 and 09-02-00664),
  the Ministry of Education, Culture, Sports, Science and Technology,
  Japan, Grant-in-Aid for Scientific Research (18071005, 19034011, 19740162),
  Swiss Nationalfonds Foundation 200020-117913/1 
  and ETH Research Grant TH-01 07-3.

\end{acknowledgments}

\bigskip 
\bibliography{bibl}

\end{document}